\documentclass[journal]{IEEEtran}
\IEEEoverridecommandlockouts
\usepackage{cite}
\usepackage{amsmath,amssymb,amsfonts}
\usepackage{algorithm}
\usepackage{algpseudocode}
\usepackage{graphicx,,epstopdf}
\usepackage{graphics}
\usepackage{graphicx}
\usepackage{epstopdf}
\usepackage{subcaption}
\usepackage{textcomp}
\usepackage{xcolor}
\usepackage{blindtext}
\usepackage{multirow}

\def\BibTeX{{\rm B\kern-.05em{\sc i\kern-.025em b}\kern-.08em
    T\kern-.1667em\lower.7ex\hbox{E}\kern-.125emX}}
\begin{document}

\title{An Autonomous Drone System with Jamming and Relative Positioning Capabilities}

\author{Nicolas Souli, Panayiotis Kolios, and Georgios Ellinas 
\thanks{Nicolas Souli, Panayiotis Kolios, and Georgios Ellinas are with the Department of Electrical and Computer Engineering and the KIOS Research and Innovation Center of Excellence, University of Cyprus, {\tt\small \{nsouli02, pkolios, gellinas\}@ucy.ac.cy}}
\thanks{This work has been supported by the European Union's Horizon 2020 research and innovation programme under grant agreement No 833611 (CARAMEL) and 739551 (KIOS CoE) and from the Government of the Republic of Cyprus through the Directorate General for European Programmes, Coordination and Development.}}

\maketitle

\begin{abstract}
As the number of unauthorized operations of Unmanned Aerial Vehicles (UAVs) is rising, the implementation of a versatile counter-drone system is becoming a necessity. In this work, we develop a drone-based counter-drone system, that employs algorithms for detecting and tracking a rogue drone, in conjunction with wireless interception capabilities to jointly jam the rogue drone while achieving self positioning for the pursuer drone. In the proposed system a software-defined-radio (SDR) is used for switching between jamming transmissions and spectrum sweeping functionalities to achieve the desired GPS disruption and self-localization, respectively. Extensive field experiments demonstrate the effectiveness of the proposed solution in a real-world environment under various parameter settings.
\end{abstract}


\section{Introduction}
Already UAV technologies have penetrated several industrial domains, and thus unsafe and illegal drone operations are becoming a major concern~\cite{park2021survey} especially in terms of critical infrastructure security. Thus, counter-drone systems must be developed to restrain illegal/unsafe drone operations. Even though high-cost anti-drone military systems have been developed to date, such solutions are not suited for civilian use; rather, the use of localized interception technology for malicious drones provides a more feasible solution for civilian applications. Already simplistic jamming solutions have been investigated to disable commercial UAVs that employ standard GPS signals to navigate \cite{Shamaei2018a, Kassas2017, Kapoor2017}. 

In the contrary, and to tackle the jamming problem, various techniques have been applied to assist with vehicle navigation, such as the inertial navigation system (INS) \cite{Morales2018b}, as well as data fusion of different measurements, for instance, from light and range sensors \cite{Maaref2018}. Such techniques, however, are not as reliable to meet the autonomous requirements, mainly due to data degradation and loss of signal \cite{Kapoor2017, Shamaei2018}. A different approach is to collect the SOPs available in the area and use them for relative positioning \cite{Maaref2018a, Raquet2007}. These signals (such as radio signals, TV signals, etc.) are prime candidates for relative positioning, as are freely available in the environment and are transmitted at high power~\cite{Morales2018b}. Such techniques, however, assumed initial knowledge of the receiver's reference position or the location of the transmitters.

This work complements previous attempts for anti-drone systems and develops a real-time counter-drone system utilizing detection, tracking, and jamming modules to counter the rogue drone's flight with high performance. In conjunction, a relative localization module is also utilized, via the use of SOPs and inertial measurements, to maintain the positioning of the friendly UAV agent when its own GPS signal is unavailable. This is achieve via the use of an SDR that can switch between the jamming and relative positioning modules. 

This work builds on our previous work in \cite{souli2020relative} which described a novel relative positioning system that localizes a UAV agent by fusing SOP information along with inertial measurements. Specifically, it significantly extends \cite{souli2020relative} as follows:
\vspace{-1pt}
\begin{itemize}
\item A novel autonomous drone system is developed that integrates both jamming capabilities as well as relative positioning technologies to achieve a two-fold objective, namely real-time rogue-drone interception and relative localization of the pursuer drone cosidering that its GPS received signal become unavailable due to the jamming operation. In particular, this integrated system:
\begin{itemize}
\item Employs an algorithm that detects, tracks and wireless jams a rogue drone in real-time.
\item Utilizes an SDR for both jamming rogue drones as well as self relatively localization via the fusion of SOP information along with inertial measurements. 
\end{itemize}
\item To assess and validate the system's performance, extensive experimental results are utilized for a wide variety of parameters in a real-world outdoor environment.
\end{itemize}

The rest of the paper is organized as follows. Section~\ref{related} presents the related work, while the methodology and the framework of the proposed technique are discussed in Sections~\ref{methodology} and~\ref{framework}, respectively. Performance results are presented in Section~\ref{results} and the main conclusions are discussed in Section~\ref{conclusions}, including directions for future research.

\section{Related Work}\label{related}
The extensive civilian use of UAVs and the noncompliance of a large number of drone users with the relevant drone operation legislations has spurred the need for effective anti-drone systems \cite{vcisar2020principles, shi2018anti}. A number of these systems are SDR-based; for example, in \cite{fang2018design}, the authors focused on a UAV GPS interference framework in low-altitude airspace  in order to induce virtual location and UAV communication signal suppression. Similarly, in \cite{brito2019jamming}, the authors developed jamming techniques using low-cost SDR technology, successfully demonstrating GPS jamming, spoofing, and remote control interference, via experimental results. Further, an SDR-based UAV GPS jamming approach is described in \cite{rahman2021unmanned}, where the performance of the jamming signal and the elevation angle on the jamming attack's range were studied. Moreover, effective GPS jamming techniques for UAVs using low-cost SDR platforms were also presented in \cite{ferreira2020effective}, demonstrating navigation interception in a number of real-life scenarios. A similar realization was described in \cite{gaspar2019anti}, where a Lidar, accelerometer, magnetometer, and GPS receiver were employed to estimate the position of a rogue UAV, integrated with SDR platforms for jamming and spoofing.

In terms of other technologies considered, in \cite{abunada2020design} an RF based anti-drone mechanism is described, for jamming the communication link between the rogue drone and the remote controller. Also, in \cite{shi2018anti}, an anti-drone system is developed, combining passive surveillance, localization, and frequency jamming technologies. However, limitations were observed in terms of drone detection, high energy consumption, and increased computational load.

There have also been numerous efforts in the state-of-the-art concerning vehicle navigation utilizing SOPs, inertial measurements, and optical flow information. In terms of SOPs, significant literature exists concerning relative positioning techniques via the usage of SOPs \cite{Morales2018b, Maaref2018a, Raquet2007}. In general, though, such techniques operate with the assumption that the reference position of the receiver and the transmitters' locations are known a priori. In addition, other relative positioning techniques focus on inertial measurements, and optical flow data. For example, in \cite{xu2018slam}, a SLAM methodology is developed on mobile robots for reliable indoor localization, using LIDAR and RGD-B camera information. Also, in \cite{shen2016optical}, to avoid the inertial navigation system drift, an EKF-based scheme is proposed combining optical flow and inertial measurements. Further, in \cite{wang2020intelligent}, curb and vertical features are employed for localization, utilizing a multi-layer LIDAR, demonstrating that the fusion of multi-layer data using an IMU led to an accurate position estimate. Also, in \cite{brossard2020ai}, a navigation technique that calculates the current position of a vehicle by using a previously determined position utilizing only IMU information is presented, relying on Kalman filtering and deep neural networks. Finally, our work in \cite{souli2021relative}, proposed an effective real-time relative positioning technique, integrating SOPs, inertial, and optical flow modalities. 

Contrary to the research efforts detailed above, the proposed counter-drone system in this work integrates both counter-drone and relative positioning functionalities to achieve the two-fold objective of real-time rogue-drone interception and relative localization of the pursuer drone that looses its GPS signal while performing wireless interception.

\section{Integrated Detecting-Tracking-Jamming}\label{methodology}
\subsection{System Model}
The system model comprises of a pursuer UAV agent that employs an onboard camera to detect and track the rogue drone, as well as a software-defined radio (SDR) platform with the capability to switch between jamming and relative positioning functionalities. Without loss of generality, jamming is enabled at the most commonly used GPS band (band L1 at $1575$MHz), to intercept the detected rogue drone's GPS. Localization is performed via SOPs and inertia measurement information collected using an onboard unit.  

\subsection{Detecting-and-Tracking Methodology}
A vision-based algorithm is utilized for detecting-and-tracking the rogue drone in the vicinity. Initially, footage from various UAV sightings were acquired and applied as a training dataset that resulted in approximately $10000$ labeled images. To implement the detection module, an onboard unit processes the captured video frames received by the pursuer UAV's camera using a convolutional neural network object detection algorithm, using the Tiny Yolov4 \cite{yolov4} framework that support real-time applications with high accuracy performance. Subsequently, to retain the rogue drone's trajectory over time, a tracker algorithm is employed, using the intersection-of-union scores method and the Hungarian algorithm as discussed in \cite{harpypaper}. Then the Euclidean distance is computed $\mathbf{d_c}$ between the pursuer and rogue UAVs as follows. Initially the focal length is calculated using:
\vspace{-2pt}
\begin{equation}
\mathbf{f}=\frac{w_{A^p}*d_r}{w_{A^m}}
\label{focal}
\end{equation}
\vspace{-2pt}
\noindent where $d_r$ is the reference distance of acquisition and $w_{A^m}$ and $w_{A^p}$ the reference object's width, in meters and pixels respectively. Subsequently, $\mathbf{d_c}$ can be obtained from:
\vspace{-2pt}
\begin{equation}
\mathbf{d_c}=\frac{w_{R^m}*\mathbf{f}}{w_{R^p}}.
\label{euclidean}
\end{equation}
\vspace{-2pt}
\noindent with $w_{R^m}$ and $w_{R^p}$ denoting the width of the detected rogue drone in meters and pixels, respectively.

\subsection{Jamming Signal Model}
Evidently, the physical layer of wireless systems can be easily targeted by jamming attacks, aiming to disrupt the communication link between communicating stations including the ground controller and a UAV, and vital UAV subsystems including GPS. In the later case, and in the absence of the jamming attack, the GPS signal is denoted as:
\vspace{-2pt}
\begin{equation} 
y_{GPS}(t)=x_{GPS}(t)+n_{GPS}(t),
\end{equation}
\vspace{-2pt}
\noindent where $x_{GPS}(t)$ is the desired signal, $y_{GPS}(t)$ is the received signal, and $n_{GPS}(t)$ is the additive Gaussian noise at time $t$. When the jamming signal is active, occupying the L1 band, the received signal becomes 
\vspace{-2pt}
\begin{equation}
y _{GPS}(t)=x _{GPS}(t)+x_{jam}(t)+n _{GPS}(t)+n_{jam}(t),
\end{equation}
\vspace{-2pt}
\noindent where $x_{jam}(t)$ is the jammer's signal and $n_{jam}(t)$ the noise between the jammer location and the receiver. In jamming attack scenarios, the received signal strength (RSS) at the receiver decreases and can be expressed as 
\vspace{-2pt}
\begin{equation}
RSS=\frac{P_{t} * G_{t} * G_{r} *\left(h t^{2} * h r^{2}\right)}{d_{rt}^{4}},
\end{equation}
\vspace{-2pt}
\noindent where $P_{t}$ is the transmitter signal power, $G_{t}$ and $G_{r}$ are the transmitter and receiver antenna gains, respectively, $h t$ and $h r$ are the transmitter and receiver antenna heights, respectively, and $d_{rt}$ is the transmitter-receiver link.

\subsection{Relative Positioning Methodology}
Relative positioning can be achieved using SOPs transmitted from a number of fixed transmitters $(T_{i}), i\in\mathcal{I}$. Using SOP measurements, vehicle positioning can be estimated using a state vector of planar information $x_{T_i}$, $p_{T_i} =[p(1)\quad p(2)]^{^{T}}$ with discrete model $p_{T_i}^{n+1}= \mathbf{F}p_{T_i}^{n}+\mathbf{w}$, where $\mathbf{F}$ designates the dynamics of the system and $\mathbf{w}$ the process noise. 

Specifically, the path-loss (P-L) signal propagation model is used to evaluate the distance ($d(i,j)$) between the fixed $i$-th ($i\in\mathcal{I}$) estimated SOP transmitter, ($p_{T_i}\in\mathbb{R}^2$) and the unknown location of the SDR receiver ($p_{R_j}\in \mathbb{R}^2$) mounted on the UAV. Subsequently, $d(i,j)$ is utilized to describe in 2D space the position of the receiver, \cite{Tomic2018}. The position estimate can be achieved using a conventional multilateration method with at least four transmitters. Computing the distances $d(i,j) \quad \forall i \in \mathcal{I}$ leads to an overdetermined system that does not have a single unique solution. An approximated solution $\mathbf{p}^{*}$ to the problem can be achieved by utilizing a LSQ technique \cite{Brown2011} after the system's linearization. 

Hence an extended Kalman filter (EKF) can be employed to produce position estimates ($\mathbf{\hat{p}}$) by fusing the onboard unit's inertial measurements with the SOPs' information, as elaborated in \cite{souli2020relative, souli2021relative}. Specifically, EKF can predict the vehicle's position at time $n+1$, ($\mathbf{\hat{p}}^{n+1}_{+}$), and create an estimated trajectory, utilizing the inertial data for the vehicle's state at time $n$ \cite{Brown2011, Sazdovski2005}. 

\section{RPS-Jamming Algorithm}\label{framework}
\begin{figure}[h]
\begin{center}
\includegraphics[width=0.9\columnwidth]{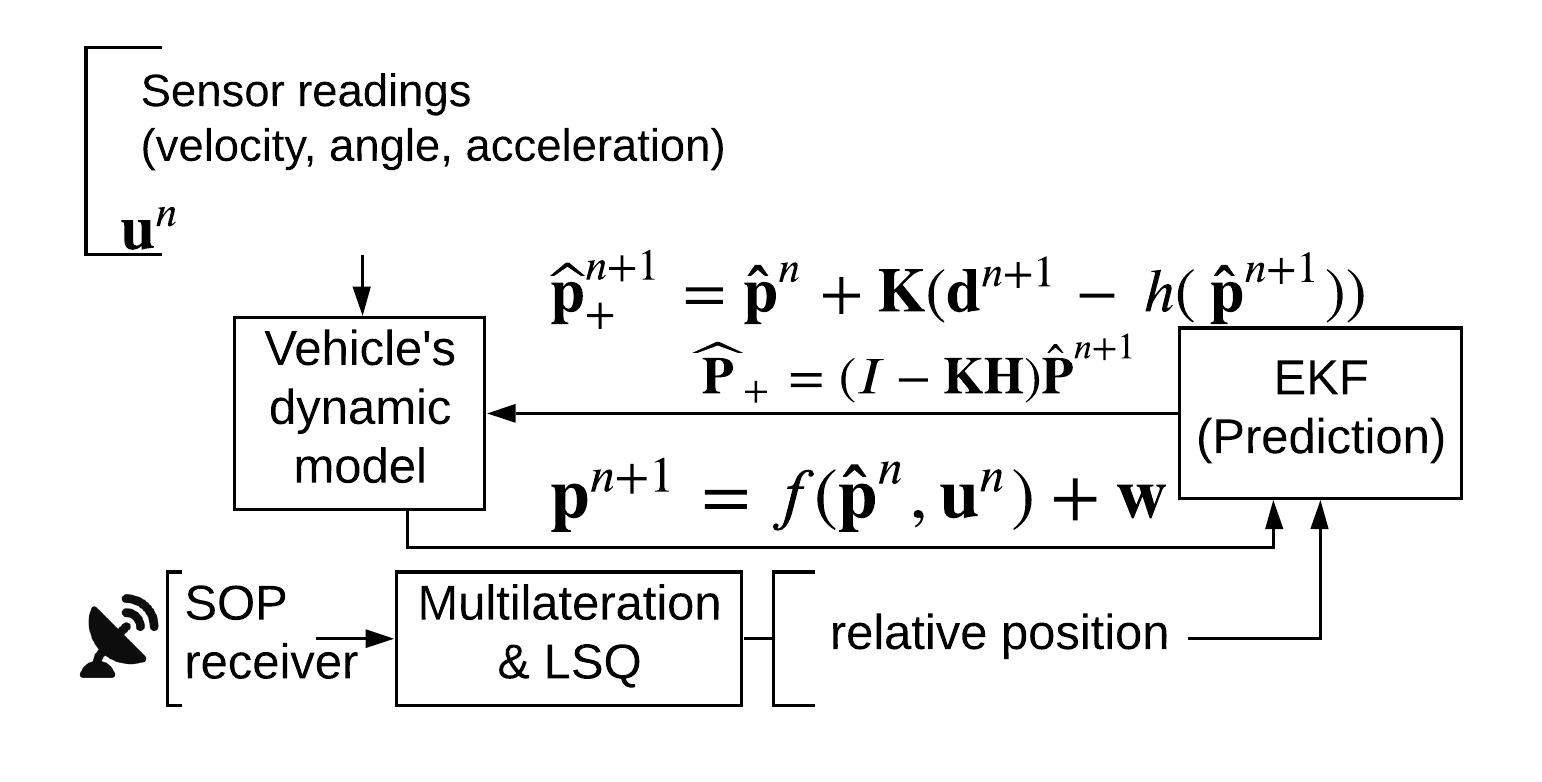}
\caption{RPS architecture.}
    \label{fig:EKF}
\end{center}
\end{figure}

This section elaborates on the proposed integrated relative positioning and jamming (RPS-JS) algorithm for detecting, tracking and jamming a rogue drone while maintaining positioning by the pursuer agent as summarized in Alg.~\ref{alg:coordinates2}. Initially a total of $\mathbf{X}^{n}$ frequency sweeps for a set of frequency bands $m$ are considered, and for each one of them the mean $\psi ^{n}_{m}$ is found. Subsequently, $\psi ^{n}_{m}$ is used in conjunction with the path-loss model, multilateration, LSQ, and EKF techniques (detailed in the previous section) to find the pursuer agents relative coordinates and obtain $\mathbf{\hat{p}}^{n+1}$. A block diagram of the RPS process flow that uses the received signal strength (RSS) of the SOP along with the multilateration, LSQ, and EKF procedures is illustrated in Fig.~\ref{fig:EKF} \cite{souli2020relative}, where $\mathbf{p}^{n}=[p_{(1)} \; \;p_{(2)}]^{T}$ gives the current planar vehicle position, $\mathbf{p}^{n+1}=f(\mathbf{\hat{p}}^{n},\mathbf{u}^{n})$ is the next estimated position, and the relative coordinates can be computed as follows:
\vspace{-2pt}
\begin{equation}
\mathbf{\hat{p}}^{n+1}=\mathbf{\hat{p}}^{n}+\mathbf{K}(\mathbf{d}^{n+1}-h(\mathbf{p}^{n}))
\label{xk+12}
\end{equation}
\vspace{-2pt}
\noindent where the error covariance $\widehat{\mathbf{P}}^{n+1}$, the Kalman gain ($\mathbf{K}$) and the updated covariance matrix error ($\mathbf{\widehat{P}_{+}}$) are defined as:
\vspace{-2pt}
\begin{eqnarray}
\widehat{\mathbf{P}}^{n+1} = \mathbf{F}\widehat{\mathbf{P}}^{n}\mathbf{F}^{\top}+\mathbf{Q}\\
\label{phat}
\mathbf{K}=\widehat{\mathbf{P}}^{n+1}\mathbf{H^{\top}}(\mathbf{H}\widehat{\mathbf{P}}^{n+1}\mathbf{H^{\top}}+\mathbf{R})^{-1}\\
\label{K}
\mathbf{\widehat{P}}_{+}=(I-\mathbf{K}\mathbf{H})\widehat{\mathbf{P}}^{n+1}
\label{widehatP}
\end{eqnarray}
\vspace{-4pt}

In addition and in order to detect-and-track the rogue drone the distance between the pursuer and rogue UAVs is calculated (i.e., $\mathbf{d_c}$). If the distance is below a predefined threshold $d_{jam}$ (i.e., the rogue drone is in range), then jamming can effectively be conducted up until $e_{m}>T_{d}$, where $e_{m}$ is the position error (in meters) for the pursuer agent and $T_{d}$ is defined as the switching threshold value, computed as the added a-th percentile (of the mean range measurement) to the mean value of the RPS relative range measurements at time $n$, $\mathbf{\bar{d}}^{n}$, prior to the jamming attack. In this way both accurate positioning and effective jamming can be achieved. 
\begin{algorithm}[h]
\caption{RPS-JS Algorithm}\label{alg:coordinates2}
\hspace*{\algorithmicindent} \small {\textbf{Input:} {RSS data obtained at each point in the route ($\mathbf{X}^{n}$)}}
\begin{algorithmic}[1]
\State \small {Scan $\mathbf{X}^{n}$ at vehicle's current position and set $T_{i}$}
\While {$n<N$} 
\State \small {{\bf Execute Relative Position Algorithm (RPS)}}
\State \small {Set the current truth value utilizing $\hat{{\mathbf{p}}}$}
\Procedure \small {{Vision data extraction}}{}
\State \small {Detect and track the UAV(s)}
\State \small {Calculate depth distance and focal length}
\State \small {Compute horizontal distance} 
\State \small {Find average of distances (for depth/horizontal)} 
\EndProcedure
\State \small {Analyze $\mathbf{X}^{n}$ and extract the distribution moments $\psi^{n}_{m}$} 
\State \small {Use $\psi^{n}_{m}$ to compute relative distances via P-L model} 
\State \small {Apply multilateration}
\State \small {Perform LSQ}
\State \small {Estimate relative position ${{\mathbf{\hat{p}}}^{n+1}}$ using EKF}
\Procedure \small {{Jamming Attack}}{}
\State \small {If $\mathbf{d_c} \leq d_{jam}$ jamming starts (and RPS is paused)}  
\State \hspace{4pt} \small {While $e_{m}<T_{d}$ Continue jamming}
\State \hspace{10pt} \small {Else Goto line 3 (RPS restarts)}
\EndProcedure
\EndWhile
\end{algorithmic}
\end{algorithm}

\begin{figure}[h]
\begin{center}
\includegraphics[width=0.9\linewidth]{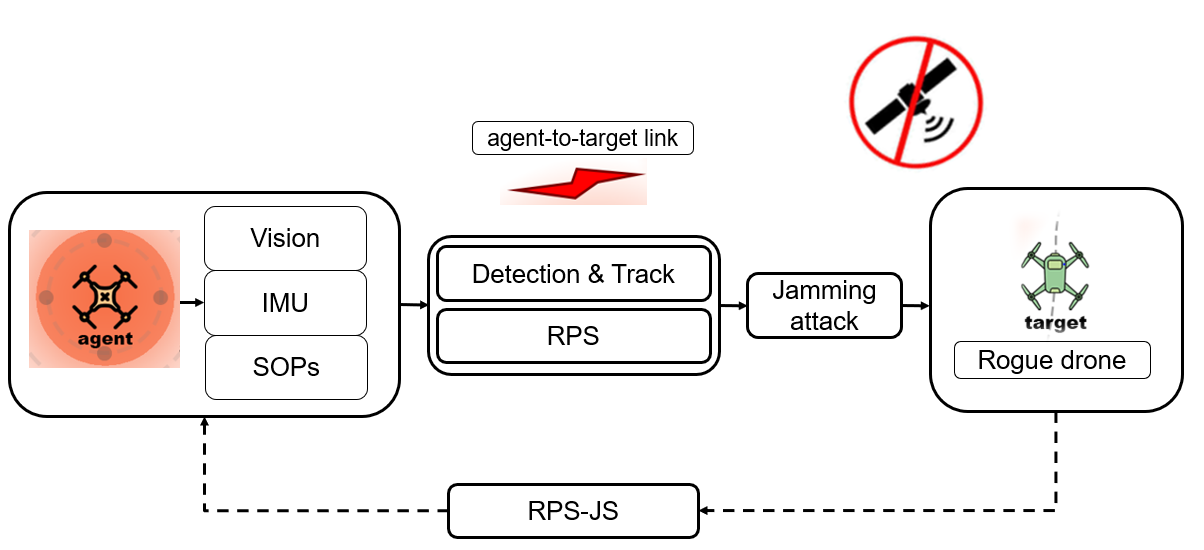}
\caption{RPS-JS algorithm's block diagram.}
    \label{fig:RPSBD2}
\end{center}
\end{figure}

The RPS-JS algorithm's block diagram is also shown in Fig.~\ref{fig:RPSBD2}, illustrating the complete operation of the system.

\section{Experimental Results}\label{results}
In order to evaluate the proposed counter-drone system, several experiments were performed, by employing the aforementioned algorithm. Various parameter settings were also considered to examine the switching between the relative positioning and jamming procedures.

\subsection{Hardware and Software Configuration}
\label{hardware-software}
\begin{figure}[h]
\begin{center}
 \setlength\abovecaptionskip{-0\baselineskip}
 \setlength\belowcaptionskip{-0pt}
\includegraphics[width=8.5cm]{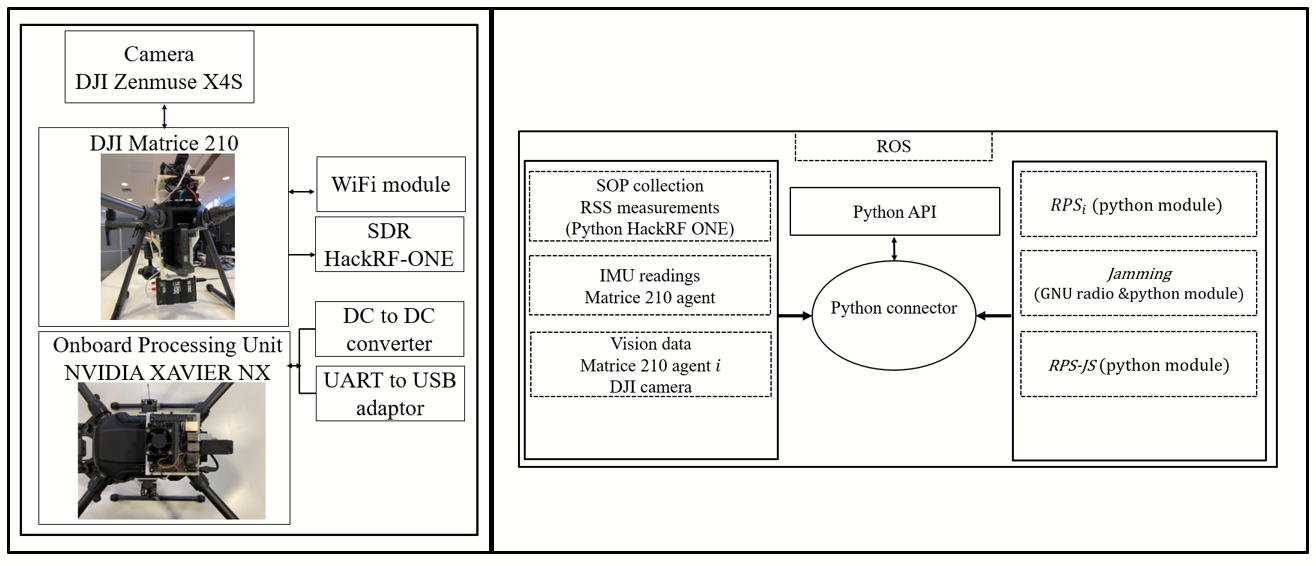}
\caption{Hardware and software implementation.}
    \label{fig:UAVhardware}
\end{center}
\end{figure}

Figure~\ref{fig:UAVhardware} depicts the proposed system hardware/software implementation. An onboard processing unit, NVIDIA Jetson Nano Developer Kit, serves as the processing unit that executes all the algorithm procedures. It is also responsible for the communication between the pursuer UAV, SDR hardware interfaces, and the remote controller application. For the communication between the application and the UAV, a WiFi dongle is employed to enable the transmission of data and commands. Further, a HackRF SDR is used for the collection of SOP measurements and the implementation of the jamming procedure. Finally, the Robot Operating System (ROS)~\cite{ROS} has been utilized, allowing command and control of the pursuer agent (i.e., receiving telemetry data and transmitting trajectory commands).

\vspace{-0.2in}

\subsection{Jamming Performance Analysis}\label{jammingperf}
\begin{figure}[t!]
\centering
\includegraphics[width=6.5cm]{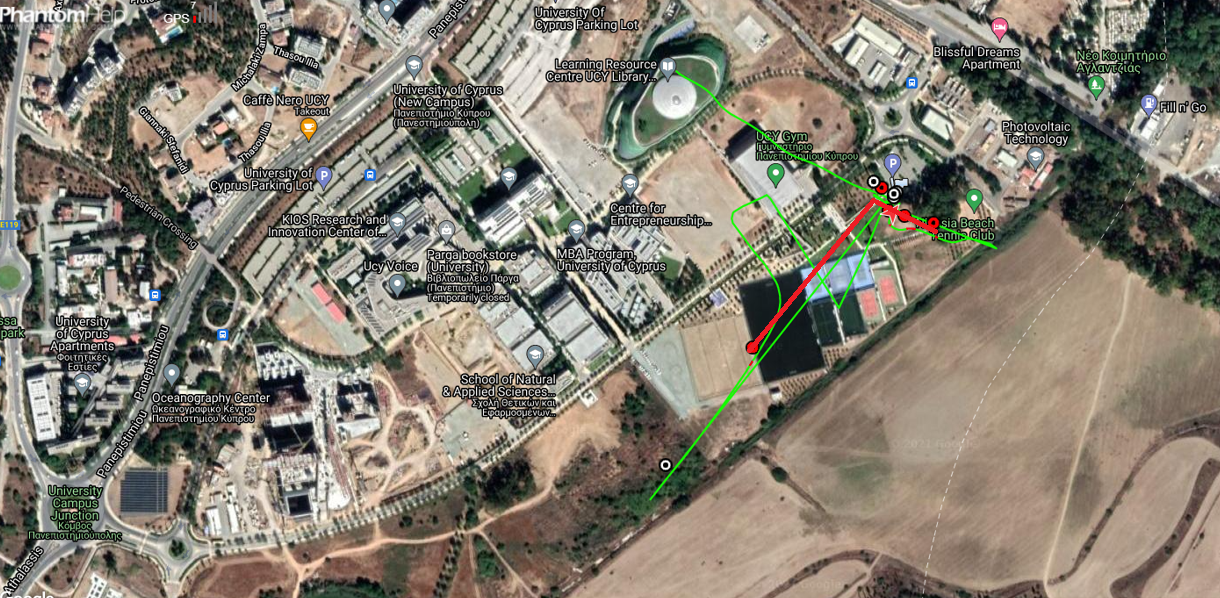}
\caption{Jamming attack on a DJI Mavic employing only GPS.}
\label{fig:GNSS}
\end{figure}

\begin{figure}[h!]
\centering
\begin{center}
 \setlength\abovecaptionskip{-0\baselineskip}
 \setlength\belowcaptionskip{-0 pt}
\includegraphics[width=9cm]{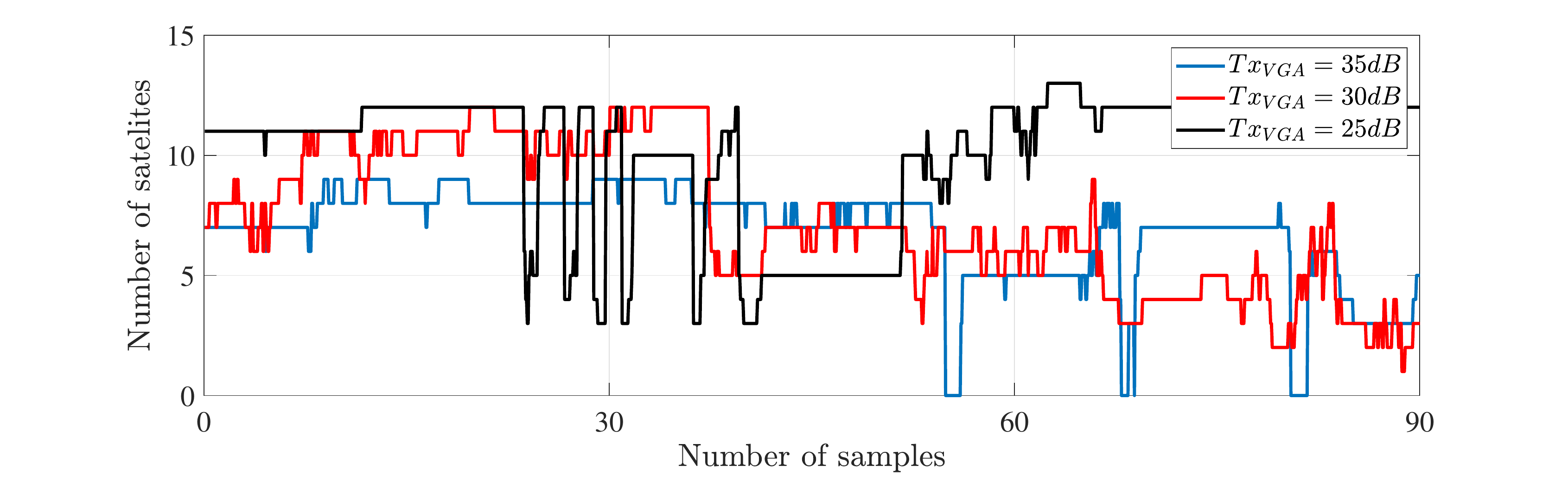}
\caption{Rogue UAV's $N_s$ values during flight (for different jamming power values).}
    \label{fig:sat_comparison}
\end{center}
\end{figure}

\begin{figure}[h]
\begin{center}
 \setlength\abovecaptionskip{0.5\baselineskip}
 \setlength\belowcaptionskip{-0 pt}
\includegraphics[width=8cm]{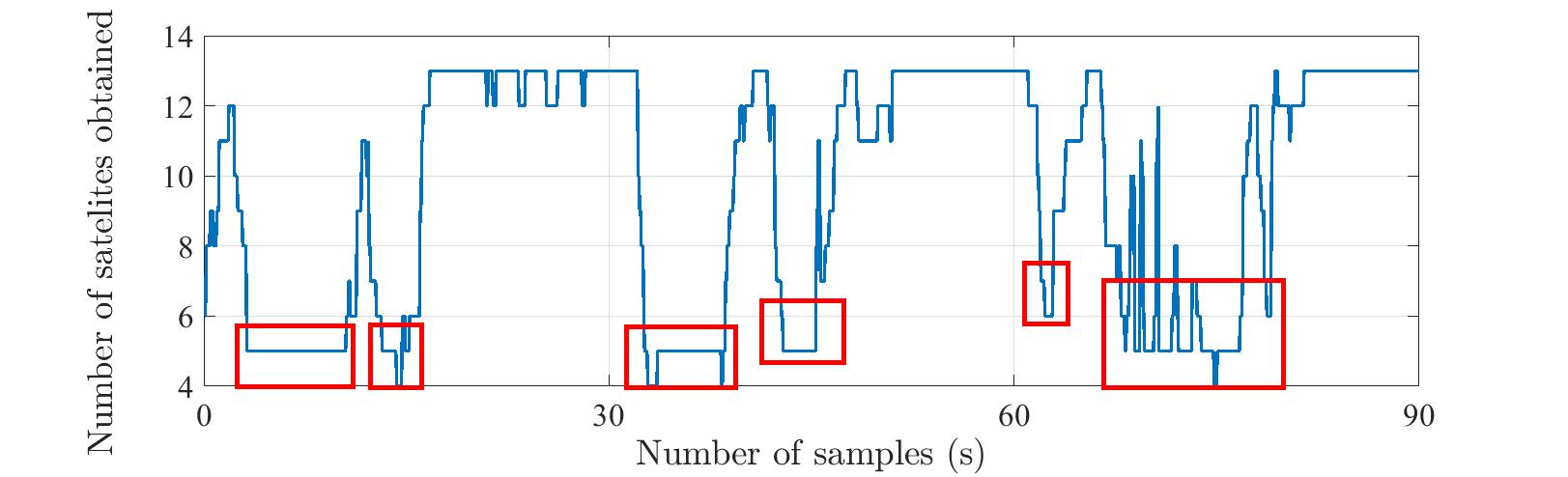}
\caption{Rogue UAV's $N_s$ values during flight (red boxes indicate jamming attacks.}
\label{fig:convergence}
\end{center}
\end{figure}

To analyze the performance of the jamming module, an omnidirectional antenna was connected to the HackRF and an initial outdoor experiment was conducted, using a DJI Mavic drone. The rogue drone's GPS is affected considerably by the implemented SDR-based jamming module (Fig.~\ref{fig:GNSS}: jammed trajectory (green) diverges significantly from the ground truth (red) shown in the figure), with the jamming module's performance evaluated for various distances and altitudes (Table I) and different transmission jamming power values. As expected, as the transmission power increases, higher jamming performance is achieved, since the number of satellites ($N_s$) obtained by the rogue drone is decreasing (Fig.~\ref{fig:sat_comparison}). However, the decreasing number of satellite signals $N_s$ due to jamming (Fig.~\ref{fig:convergence}, red boxes), also affect the pursuer agent (since we consider in this experiment the off-the-shelf GPS receivers with omnidirectional antennas). Hence, the tradeoff between jamming duration and possitioning accuracy achieved using the proposed RPS framework (applied by the pursuer agent) is addressed using the $T_{d}$ threshold as examined below.

\begin{table}[h]
\begin{center}
\caption{Outdoor Jamming Experiments}
\begin{tabular}{|l|l|l|l|l|l|l|}
\hline
\textbf{Altitude (m)}          & \multicolumn{2}{l|}{25} & \multicolumn{2}{l|}{30} & \multicolumn{2}{l|}{50} \\ \hline
\textbf{Absolute distance (m)} & 5          & 10         & 5          & 10         & 5          & 10         \\ \hline
\textbf{Result (GPS)}          & x          & x          & x          & x          & x          & -          \\ \hline
\textbf{Elevation angle}       & 0          & 30         & 0          & 30         & 0          & 30         \\ \hline
\multicolumn{7}{|l|}{\textbf{x : No GPS, - : GPS unaffected}}                                               \\ \hline
\end{tabular}
\end{center}
\label{tab:I}
\end{table}

\vspace{-0.2in}

\vspace{-0.2in}

\subsection{RPS-JS Experimentation}
The performance of the proposed counter-drone system methodology is evaluated in a number of outdoor field experiments. In these experiments, the frequency bands $0-3500$ MHz were swept for SOPs, and the RPS parameters $T_{i}=13$ and $n_{PL}=2.8$ (path loss exponent) were used, based on previous experimental results \cite{souli2020relative} that have shown fast system convergence to the relative coordinates.

\begin{figure}[h]
\begin{center}
 \setlength\abovecaptionskip{-0\baselineskip}
 \setlength\belowcaptionskip{-0pt}
\includegraphics[width=9.1cm]{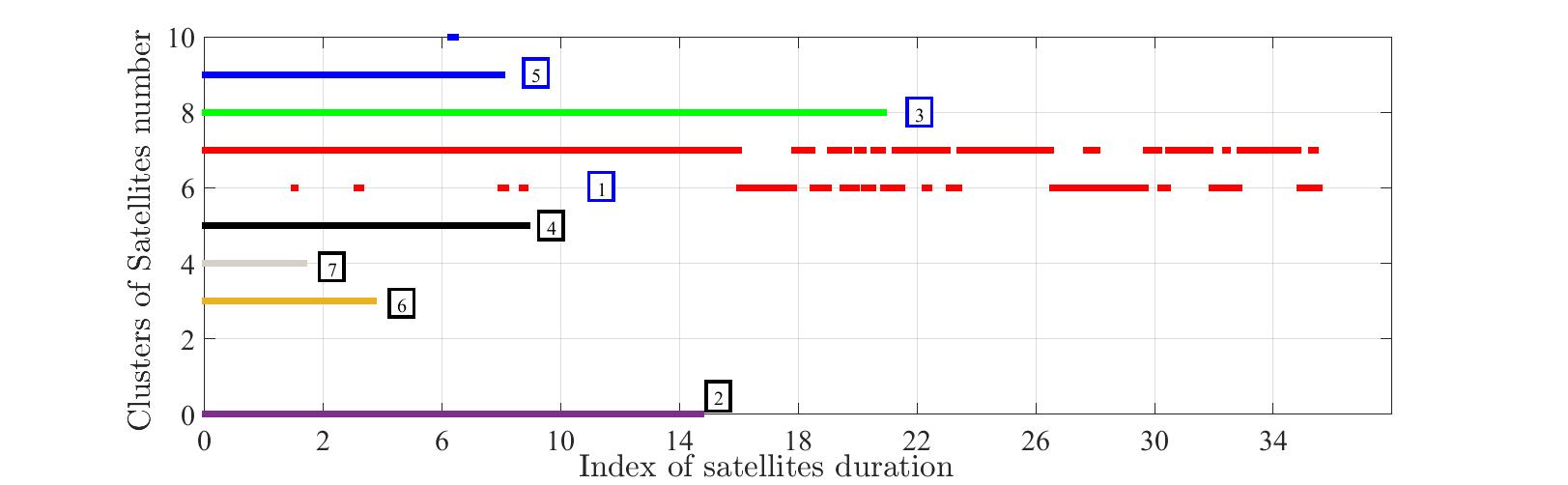}
\caption{Rogue UAV's $N_s$ values during flight (during RPS/jamming switching).} 
  \label{fig:RPSONLRPSEXECUTIONTIME}
\end{center}
\end{figure}

\begin{figure}[h]
\begin{center}
 \setlength\abovecaptionskip{-0\baselineskip}
 \setlength\belowcaptionskip{-0pt}
\includegraphics[width=9.1cm]{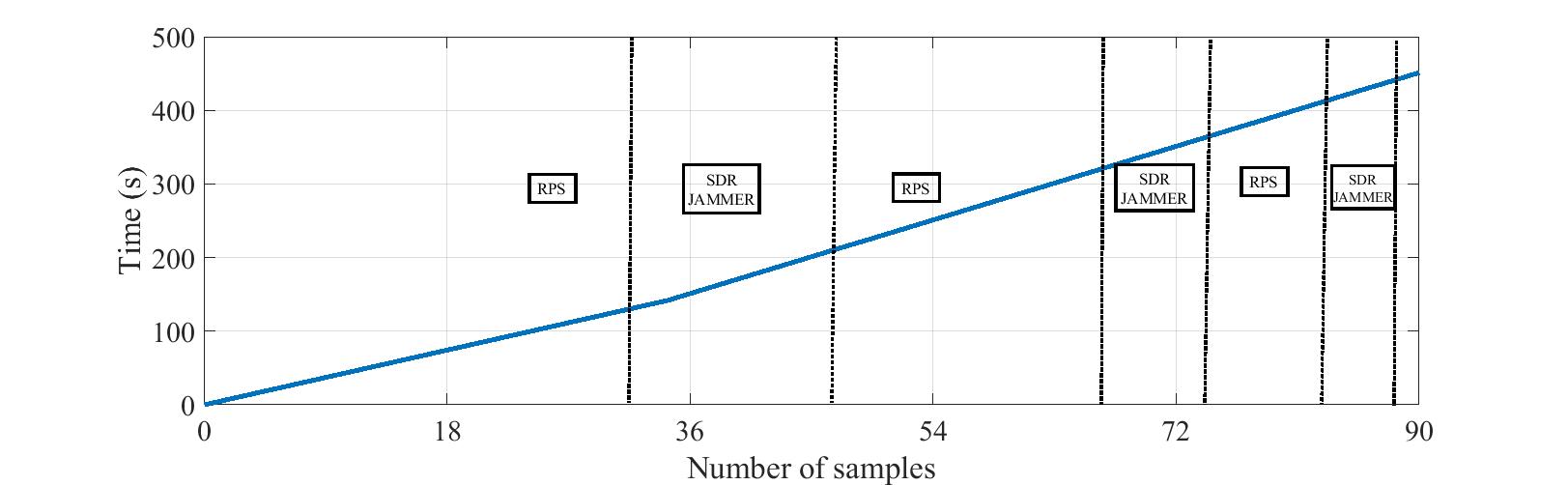}
\caption{RPS/jamming modules' execution times over flight time.}
  \label{fig:RPSjamEXECUTIONTIME}
\end{center}
\end{figure}

Initially, Fig.~\ref{fig:RPSONLRPSEXECUTIONTIME} depicts the number of satellites ($N_s$) available when the proposed system switches between positioning and jamming (shown with blue and black boxes, respectively, with the insert numbers denoting the sequence of the events), demonstrating that as the number of jamming attacks increases the rogue UAV cannot regain its location accuracy (i.e., $N_s$ decreases). Further, Fig.~\ref{fig:RPSjamEXECUTIONTIME}, illustrates the execution time for both the RPS and jamming modules over $2600$ samples ($N=2600$) and a threshold set at $T_d=1.3m$, for a complex route of $300$m and an approximate $8$ minutes of flight, showing a substantial decrease of both modules' execution time as the flight time increases. This is due to the fact that  the value of $e_{m}$ accumulates over time, due to the rapid switching between the RPS and jamming modules that need to alternate in order to maintain the relative position estimation while at the same time effectively jam the rogue drone's GPS.

As previously discussed in Section~\ref{jammingperf}, when a jamming attack occurs, the UAV agent's GPS data are not sufficient for positioning, and a position estimate for the navigation of the UAV agent that can compensate for the agent's GPS decreased accuracy is a necessity. Thus, for various routes in outdoor experiments, a comparison of range measurements between the real, relative, and GPS trajectories, measured in meters, using proposed the counter-drone system in an online fashion, is tabulated in Table II, showing the difference between the two systems in terms of localization performance. It is notable that the proposed RPS-JS system leads to accurate relative positioning even when extended jamming attacks occur.

\begin{table*}[t]
\begin{center}
\caption{Online Experiment - Trajectory Distances}
\begin{tabular}{|l|l|l|l|l|l|}
\hline
\textbf{Trajectory Segments} & \textbf{A-B(m)/Dif (\%)} & \textbf{B-C(m)/Dif(\%)} & \textbf{C-D(m)/Dif(\%)}& \textbf{D-E(m)/Dif(\%)} & \textbf{E-F(m)/Dif(\%)}\\ \hline
GT distance              & 20/0                     & 15/0                    & 30/0              & 15/0   &  13/0    \\ \hline
Relative distance - (RPS-JS)  &18/10              & 13.8/8           & 26/13      & 13/13.33    &  11/15.3        \\ \hline
GPS+sensors distance & 19/5            & 14/6.66          & 20/33.33     &10/33.33    &21/61.5    \\ \hline
\multicolumn{6}{|l|}{RPS-JS configuration  ($T_{i}=13$, $n_{PL}=2.8$, $T_{d}=1.3m$)}            \\ \hline
\end{tabular}
\end{center}
\label{tab:trajectorydistances3}
\end{table*}

\begin{figure}[h]
\centering
\begin{subfigure}[b]{\columnwidth}
\centering
\includegraphics[width=9cm]{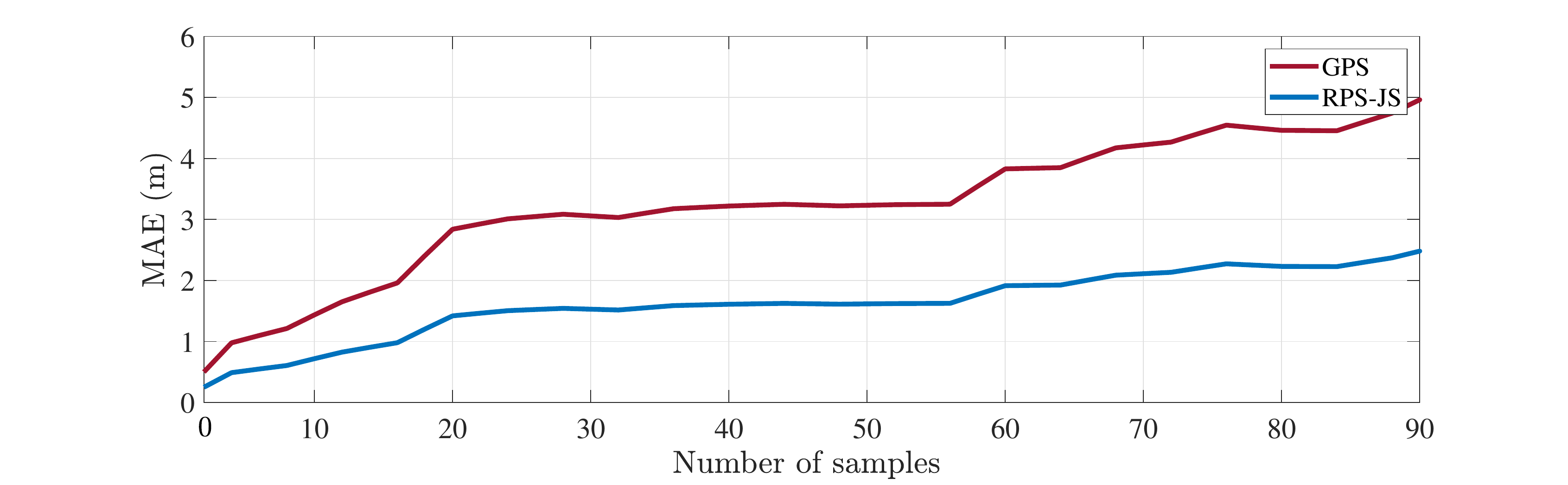}
\caption{}
    \label{}
\end{subfigure}
\begin{subfigure}[b]{\columnwidth}
\centering
\includegraphics[width=9cm]{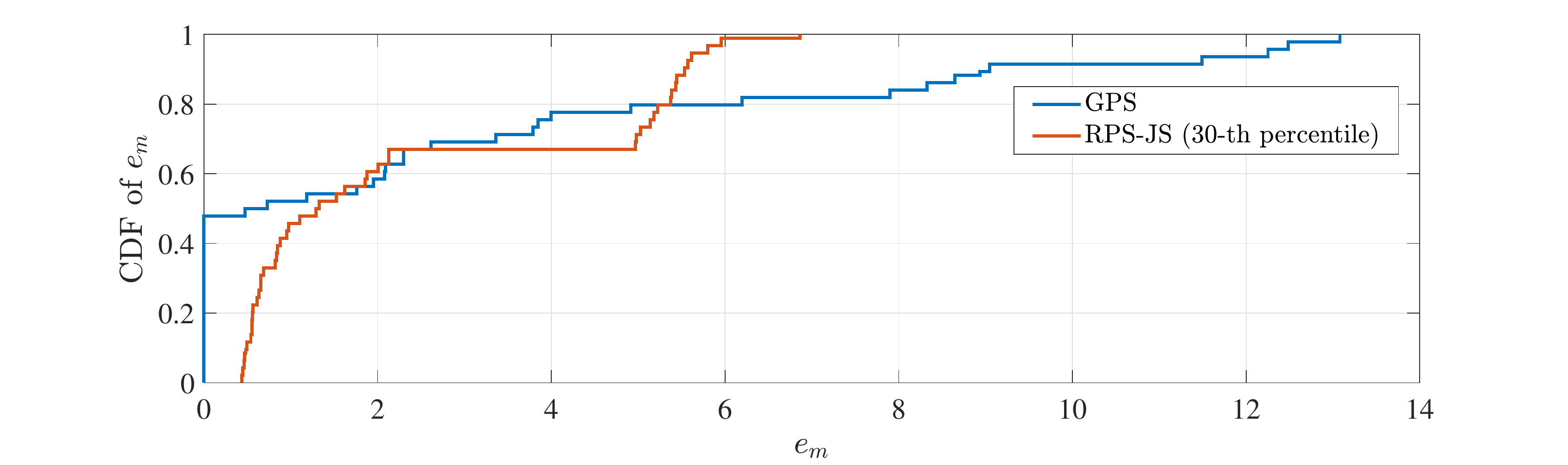}
\caption{}
\label{}
\end{subfigure}
\caption{(a) MAE (in meters) of RPS-JS and GPS vs. GT (b) CDF of position estimation error for RPS-JS and GPS, in jamming attack scenarios.}
\label{fig:MAEANDACCUMULATED}
\end{figure}

Further, the deviations between real (GT), GPS, and relative trajectories, when applying the proposed approach, are also illustrated in Fig.~\ref{fig:MAEANDACCUMULATED}. The performance of the counter-drone system, comparing relative and GPS positioning to the GT, is evident in the case of various jamming attacks that also decrease the UAV agent's GPS accuracy (Fig.~\ref{fig:MAEANDACCUMULATED}(a)), as the mean absolute error (MAE) of RPS-JS is significantly lower compared to GPS. This is also shown utilizing a cumulative distribution function (CDF) plot of the position estimation error in Fig.~\ref{fig:MAEANDACCUMULATED}(b), which illustrates that the position estimation error of RPS-JS is $90\%$ of the time less than $6$m as compared with GPS, which has a value greater than $6$m at $50\%$ of the time.

\begin{figure}[t!]
\centering
\begin{subfigure}[b]{\columnwidth}
\centering
\includegraphics[width=9cm]{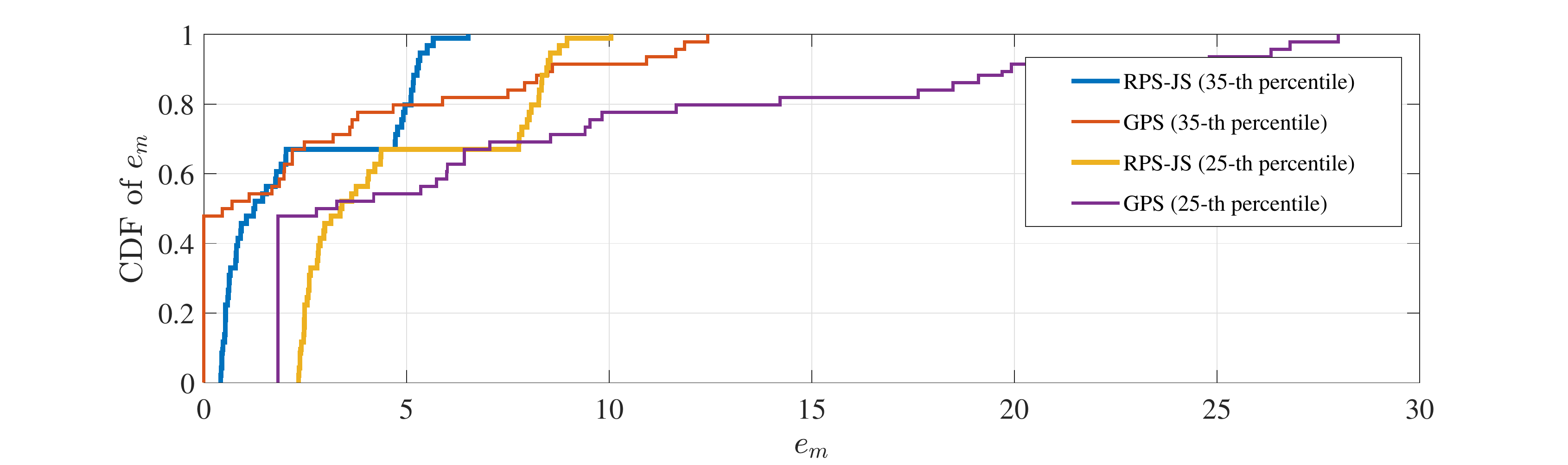}
\caption{}
\label{}
\end{subfigure}
\begin{subfigure}[b]{\columnwidth}
\centering
\includegraphics[width=9cm]{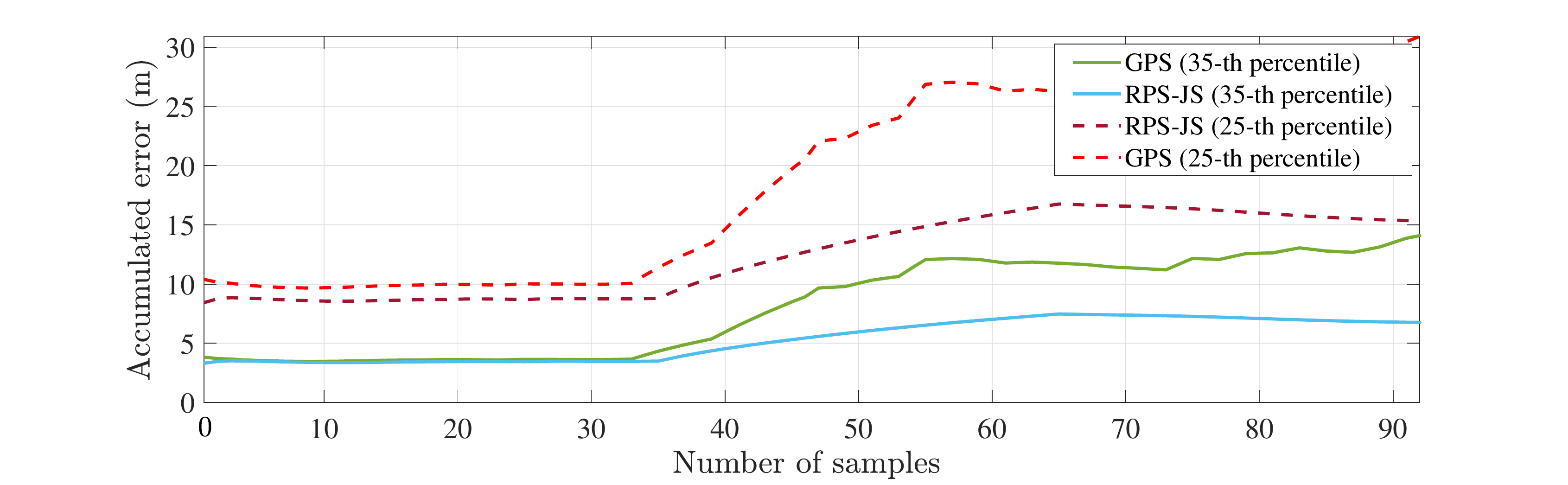}
\caption{}
\label{}
\end{subfigure}
\caption{(a) CDF of range measurements. (b) Accumulated error (in meters) using different a-th percentiles.}
\label{fig:periodicityonlrpsanalyzed}
\end{figure}

As previously discussed, RPS-JS requires a certain threshold value ($T_{d}$) in order to maintain an accurate estimation of the UAV agent while switching from jamming to relative positioning. When $e_{m}>T_{d}$ is reached, jamming is paused and the RPS process restarts so as to reduce the error and achieve the predefined $\mathbf{\bar{d}}^{n}$. Clearly, a larger $T_{d}$ will lead to a more reliable localization performance, as the switching between the jamming and relative positioning is reduced, at the expense though of the jamming performance. To illustrate this, Figure~\ref{fig:periodicityonlrpsanalyzed}(a) captures the RPS-JS and GPS behavior (MAE results) compared to the GT using various a-th percentile values, demonstrating that as the a-th percentile is reduced, switching between the two modules increases and the localization accuracy diminishes. Also, Fig.~\ref{fig:periodicityonlrpsanalyzed}(b) presents the accumulated error of RPS-JS and GPS compared to the GT by altering the a-th percentile parameter, demonstrating the trade-off between better localization performance  (when employing a larger $T_{d}$) and better jamming performance (when employing a lower $T_{d}$).

\begin{figure}[h]
\begin{center}
 \setlength\abovecaptionskip{-0\baselineskip}
 \setlength\belowcaptionskip{-0pt}
\includegraphics[width=9.1cm]{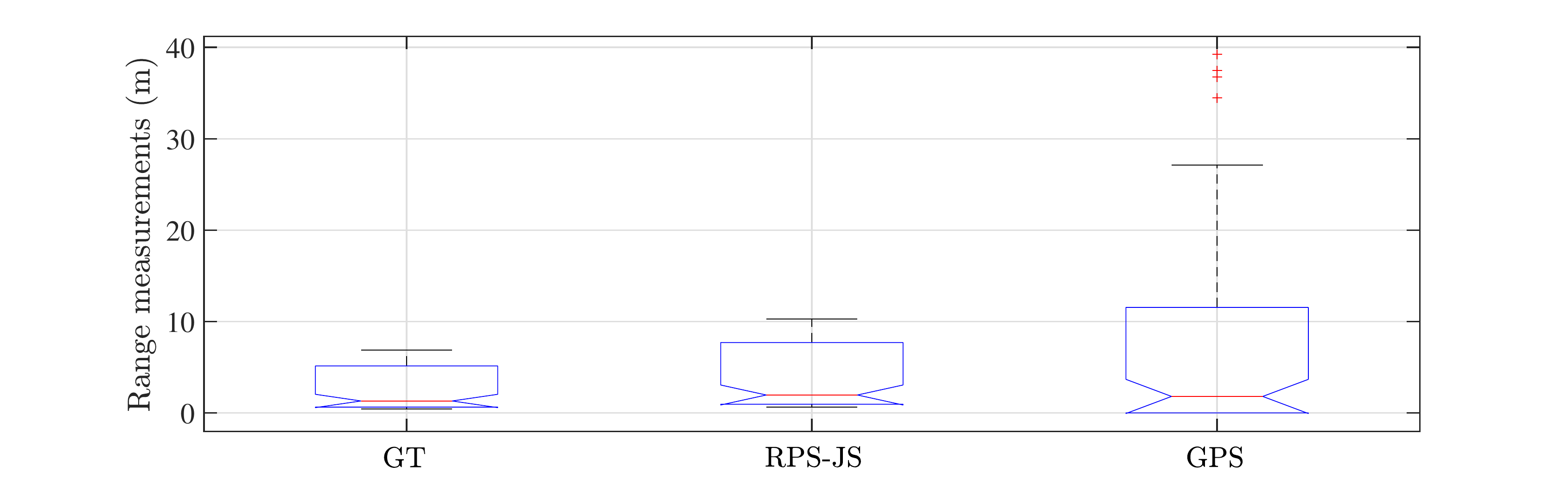}
\caption{Boxplots of RPS-JS and GPS vs. GT.}
  \label{fig:RPSjamboxplots}
\end{center}
\end{figure}

Moreover, a statistical analysis is conducted to examine the performance of RPS-JS and GPS vs. GT (Fig. ~\ref{fig:RPSjamboxplots}). It is noted that in terms of localization (in $25$ outdoor experiments), the distribution of RPS-JS achieves acceptable performance, as the median and mean values are analogous to the GT dataset.

Finally, various more complex routes were considered, and it was demonstrated that even though the complexity of the flight routes increased, the proposed system was still capable to detect, track, interfere with the positioning of a rogue drone, and self-localize. This was achieved by increasing the switching between jamming and relative positioning procedures to achieve a better localization and jamming performance. These results are omitted due to space considerations. 


\section{Conclusions}\label{conclusions}
A novel real-time counter-drone system is proposed, exploiting vision-based measurements for rogue drone detection and tracking, the information from SOPs and IMU for relative positioning, and a jamming method for hijacking a rogue drone. Field experiments, using a real hardware and software implementation of the proposed methodology, demonstrated that the proposed system can efficiently switch between the navigation module, utilizing a relative trajectory, and the jamming module, in order to increase the performance of the system (disrupt the rogue drone's actions and also navigate in case the UAV agent's GPS data are not available or unreliable).

Future research avenues include the use of a swarm of UAV agents to counter the operation of a rogue drone, as well as the use of passive radar for drone detection and for improved relative navigation via received RF information.

\bibliographystyle{IEEEtran}

\end{document}